\documentclass[a4paper]{jpconf}
\usepackage{graphicx}
\begin{document}
\title{
Self-similarity of proton spin and $\rm z$-scaling}

\author{M Tokarev$^1$, I Zborovsk\'{y}$^2$}

\address{$^1$ Joint Institute for Nuclear Research, Dubna, Moscow region,  Russia}

\address{$^2$ 
Nuclear Physics Institute,
Academy of Sciences of the Czech Republic, 
\v {R}e\v {z}, Czech Republic}

\ead{tokarev@jinr.ru}
\ead{zborovsky@ujf.cas.cz}  

\begin{abstract}
The concept of $z$-scaling previously developed for analysis of inclusive 
reactions in proton-proton collisions is applied for description 
of processes with polarized particles. Hypothesis of self-similarity
of the proton spin structure is discussed. 
The possibility of extracting information 
on spin-dependent fractal dimensions of hadrons and fragmentation 
process from the cross sections and asymmetries is justified. The double longitudinal 
spin asymmetry $A_{LL}$ of jet and $\pi^0$-meson  production and the coefficient
of polarization transfer $D_{LL}$ measured
in proton-proton collisions at $\sqrt s = 200$~GeV  
at RHIC are analyzed in the framework of $z$-scaling. 
The spin-dependent fractal dimension of proton is estimated.
\end{abstract}

\section{Introduction}
Spin is one of the most fundamental properties of elementary particles. 
The proton spin structure  is studied for a long time in processes
with polarized leptons and protons. 
The goal is to understand complete picture of the proton spin in terms
of quark and gluon degrees of freedom. 
Microscopic scenario of proton structure implies knowledge of momentum 
and spin distributions of its constituents at different scales.
In the framework of QCD, the non-linear Yang-Mills equations taking 
into account gauge invariance and Lorentz covariance regulate dynamics 
of the constituent interactions both at hard and soft regimes.

There is a convincing evidence that inclusive reactions with
unpolarized particles 
reveal self-similarity over a wide scale range \cite{1,2}. 
The scaling features of hadron production in $ p+p$ and $\bar{p}+p $
 collisions ($z$-scaling) are treated as a manifestation of self-similarity of the structure
of colliding protons, interaction mechanism 
of their constituents, and fragmentation process into real hadrons. 
The parameters of the scaling, $c$, $\delta$ and $\epsilon_F$, 
have physical interpretation of a heat capacity of the produced matter, 
fractal dimension of the proton structure and fractal dimension
of the fragmentation process, respectively.
Extension of the method for analysis of polarization phenomena 
and verification of self-similarity of particle production in polarized $p+p$  collisions 
is an interesting problem which could give new insight into our understanding 
the origin of proton spin.
The spin-dependent fractal dimensions are new parameters of the $z$-scaling approach.
They characterise polarization properties
of proton structure, constituent interactions and hadronization process.

In this paper we analyze  the double longitudinal spin asymmetry $A_{LL}$
of jet and meson  production and the coefficient of polarization transfer $D_{LL}$
measured by the STAR and PHENIX collaborations at RHIC
to extract information on spin-dependent fractal dimensions of proton
and verify the hypothesis of self-similarity of proton spin in the $z$-scaling approach.

\section{$z$-Scaling}  
The  $z$-scaling \cite{1,2} manifests itself in the fact that the inclusive spectra 
of various types of particles are described with a universal scaling function.
The function  $\Psi(z)$ depends on single variable $z$ in a wide range of the transverse momentum, 
registration angles, collision energies, and centralities. 
The scaling variable has the form:
\begin{equation}
z=z_0 \cdot \Omega^{-1}.
\label{eq:1}
\end{equation}
The quantity  $z_0$ is ratio of the transverse kinetic energy 
of the selected binary sub-process 
required for production of the inclusive particle  
with its partner (antiparticle) and the charged multiplicity density at midrapidity 
raised to a power of $c$. 
The parameter $c$ is interpreted as a heat capacity of the produced matter.
The selected binary sub-process
is given by maximum of the function 
$\Omega(x_1,x_2,y_a,y_b)= 
(1-x_1)^{\delta_1}(1-x_2)^{\delta_2}(1-y_a)^{\epsilon_a}(1-y_b)^{\epsilon_b}$ 
with the condition 
$(x_1P_1+x_2P_2-p/y_a)^2=M_X^2$,\  where $M_X=x_1M+x_2M+m/y_b$ \ is the minimal mass
of the recoil system.
The constraint expresses locality of the interaction at the constituent
level.
It restricts momentum fractions  
via kinematics of the constituent interactions 
and the 4-momenta $P_1, P_2$ and $p$ of the colliding protons and the inclusive particle with masses 
$M$ and $m$, respectively.

The function $\Omega(x_1,x_2,y_a,y_b)$  is proportional to the relative number of the configurations at the constituent level which include the binary sub-processes corresponding to the momentum fractions $x_1$ and $x_2$  of colliding protons  
and to the momentum fractions $y_a$  and $y_b$  of the secondary objects directly 
produced in the constituent interactions. 
The parameters $\delta_1$  and $\delta_2$  are fractal dimensions of the protons and 
$\epsilon_a$  and $\epsilon_b$ 
are fractal dimensions of the fragmentation process in the scattered and recoil direction, respectively. For the reactions with no polarizations we have
$\delta_1=\delta_2=\delta$ and 
$\epsilon_a=\epsilon_b=\epsilon_F$. The value of $\epsilon_F$ depends on type of the inclusive hadron. 
The parameter dependence of the maximum of   $\Omega(x_1,x_2,y_a,y_b)$   
used in (1) is briefly indicated as 
$\Omega \equiv \Omega_{0000} =: \{\delta_1,\delta_2,\epsilon_a,\epsilon_b \}$ in the next.
The low index $(0000)$ corresponds to unpolarized particles in the initial and final states. 
Similar notation applies for processes with polarizations.

    The scaling function $\Psi(z)$ is expressed in terms 
of the inclusive cross section 
$Ed^3\sigma / dp^3$, multiplicity density $ dN/d\eta$, 
and total inelastic cross section $\sigma_{in}$
for   the inclusive reaction 
$P_1+P_2\rightarrow p+X$. The function $\Psi(z)$ is determined 
by the following expression:
\begin{equation}
\Psi(z)=-{{\pi}\over {(dN/d\eta)\sigma_{in}}} J^{-1} E {{d^3\sigma}\over {dp^3} }.
\label{eq:6}
\end{equation}
Here $J$ is Jacobian for the transition from the variables $\{p_T^2,y\}$  to $\{z,\eta \}$.
The value of $\Psi(z)$ is  probability 
density to produce the inclusive particle 
with the corresponding value of the self-similarity parameter $z$.

\section{Self-similarity for processes with polarizations} 
Below we use the $z$-scaling approach to formulate self-similarity hypothesis for 
reactions with polarized particles and discuss a possibility 
for studying inclusive particle production in polarized $p+p$  collisions
to extract information on spin-dependent fractal dimensions of proton \cite{4,5}.

The reaction  $\overrightarrow{\!\!p}+\overrightarrow{\!\!p}\rightarrow h+X$  
with two longitudinally polarized protons in the initial state is described 
by the spin-dependent cross sections 
$\sigma_{++},\sigma_{--},\sigma_{+-},\sigma_{-+}$. 
The signs $(+)$ and $(-)$ denote positive and negative helicities of the protons, respectively. 
The double spin asymmetry $A_{LL}$  of the process is expressed via combination of the cross sections in the form:
\begin{equation}
A_{LL}= {   {\sigma_{++}+\sigma_{--} - \sigma_{+-} - \sigma_{-+} } \over 
{\sigma_{++}+\sigma_{--} + \sigma_{+-} + \sigma_{-+} }}.
\label{eq:11}
\end{equation}
The corresponding quantities $\Omega$   
are written as follows:
$\Omega_{++00} =: \{ \delta-\Delta \delta/2, \delta-\Delta \delta/2, \epsilon_F, \epsilon_F \}$, \ \ 
$\Omega_{--00} =: \{ \delta-\Delta \delta/2, \delta-\Delta \delta/2, \epsilon_F, \epsilon_F \} $, \ \
$\Omega_{+-00} =: \{ \delta, \delta+\Delta \delta, \epsilon_F, \epsilon_F \} $, \ 
$\Omega_{-+00} =: \{ \delta+\Delta \delta, \delta, \epsilon_F, \epsilon_F \}.$

The process $\overrightarrow{\!\!p}+p \rightarrow \overrightarrow{\!\!h}+X$ 
with one longitudinally polarized proton in the initial state and one
longitudinally polarized particle (e.g. $\Lambda$ hyperon) in the final state
is described by the coefficient of  polarization transfer. 
It is written in the following form:
\begin{equation}
D_{LL}= {   {\sigma_{++}+\sigma_{--} - \sigma_{+-} - \sigma_{-+} } \over 
{\sigma_{++}+\sigma_{--} + \sigma_{+-} + \sigma_{-+} }}.
\label{eq:16}
\end{equation}
The symbols with $(+)$ and $(-)$ denote cross sections corresponding to the parallel 
and antiparallel spin orientations relative to the respective momenta
of the polarized particles (positive and negative helicities). 
The polarization in the initial state is related to the spin-dependent correction 
($ \Delta\delta$) of the proton fractal dimension. 
Let us consider $(+)$ helicity in the initial state only and denote $\epsilon_F$  
as the fractal dimension for hadronization of an unpolarized particle ($h$)
in the final state. This corresponds to 
$\Omega_{+000} =: \{ \delta-\Delta \delta/4, \delta+\Delta \delta/4, \epsilon_F, \epsilon_F \}$.
If the inclusive particle is polarized, ($\overrightarrow{\!\!h}$),
the spin-dependent correction $\Delta \epsilon_F$  to the value of $\epsilon_F$ is included. 
The notations for such process are as follows:
$\Omega_{+0+0} =: \{ \delta-\Delta \delta/4, \delta+\Delta \delta/4,
 \epsilon_F-\Delta \epsilon_F/2, \epsilon_F \},$ \ \
$\Omega_{+0-0} =: \{ \delta-\Delta \delta/4, \delta+\Delta \delta/4,
 \epsilon_F+\Delta \epsilon_F/2, \epsilon_F \}$.

  Using experimental information on the spin asymmetries and the unpolarized cross section, 
  the spin-dependent functions  
  $\Psi_{++},\Psi_{--},\Psi_{+-},\Psi_{-+}$
  can be constructed \cite{4,5}. The functions have different arguments which we denote as
  $z_{++}, z_{--}, z_{+-}, z_{-+}$, respectively. 
  They are expressed (\ref{eq:1})  by spin-dependent fractal dimensions. Based on existence 
  of the $z$-scaling in unpolarized $p+p$ collisions,
  we consider self-similarity of polarization processes 
  at a constituent level in the form:
\begin{equation}
  \Psi_{++}=\Psi(z_{++}), \ \
  \Psi_{+-}=\Psi(z_{+-}), \ \
  \Psi_{00}=\Psi(z_{00}).
\label{eq:19}
\end{equation}
This hypothesis assumes universality of $\Psi(z)$ for different spin orientations.
The relations include corrections $\Delta \delta$  and $\Delta \epsilon_F$
to the fractal dimensions $\delta$  and $\epsilon_F$ found for unpolarized reactions.
Information on both the polarized and unpolarized cross sections are necessary 
to extract the spin-dependent fractal dimensions from the polarization 
characteristics $(A_{LL}, D_{LL})$ 
of a given process. 
Such data allows us to obtain restrictions 
on the parameters  $\Delta \delta$  and  $\Delta \epsilon_F$ of the model.

\section{$z$-Scaling and polarized $p+p$ collisions at RHIC} 
Figure 1 shows the scaling function $\Psi(z)$ for unpolarized $p+p$ collisions.
The $z$-presentation of spectra of jets measured  
at $\sqrt s = 38-200$~GeV and $\pi^0$ mesons
measured at $\sqrt s = 23-200$~GeV and $\theta_{cms}=90^0$ 
is depicted in Figs. 1(a) and 1(b), respectively.
The $z$-scaling of other hadrons is illustrated in Fig. 1(c). 
The data were obtained at the accelerators  ISR, S$\rm p\bar p$S and RHIC.
\begin{figure}[b]
\includegraphics[width=10pc,height=10pc]{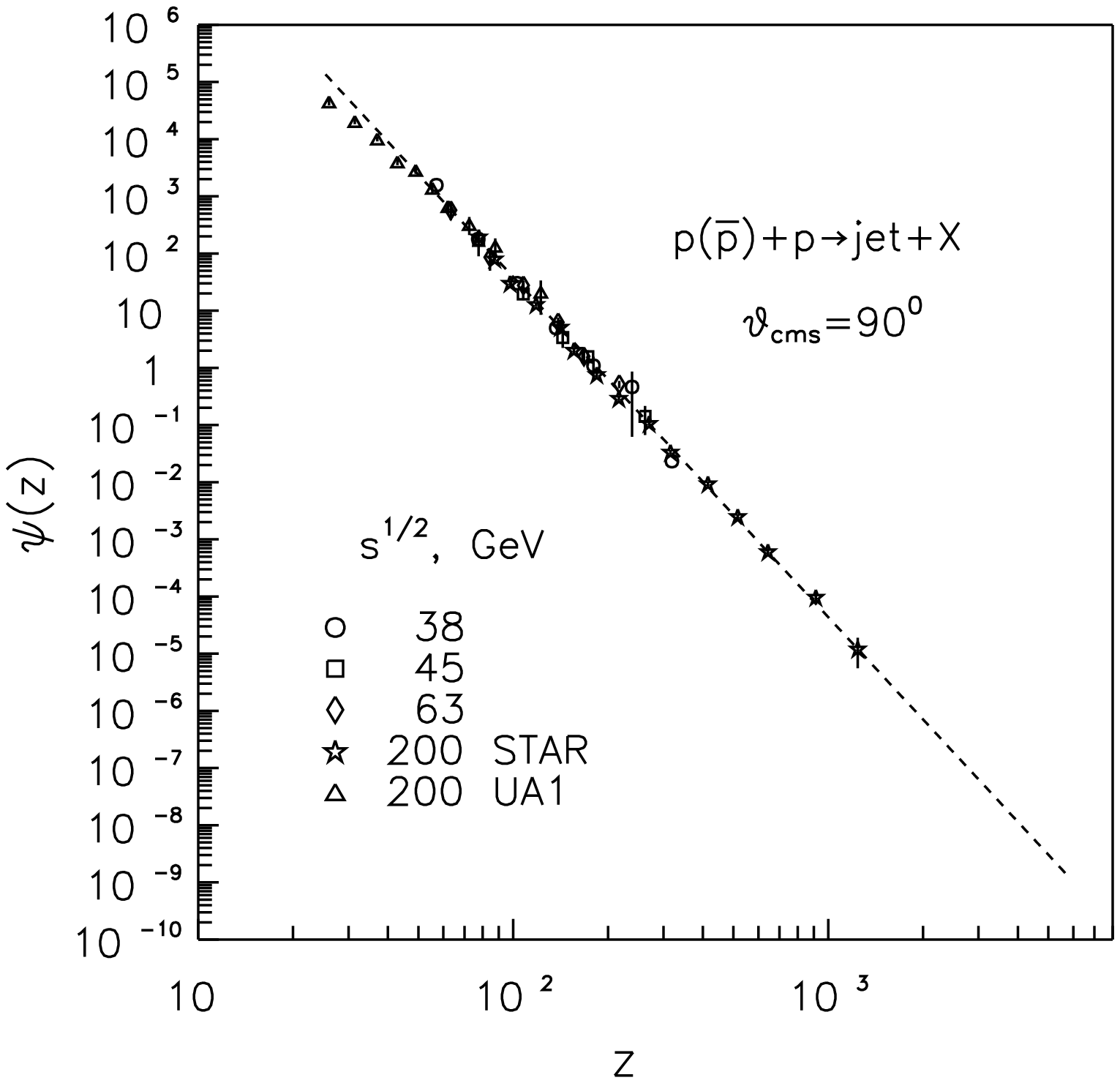}
\hspace*{5mm}
\includegraphics[width=10pc,height=10pc]{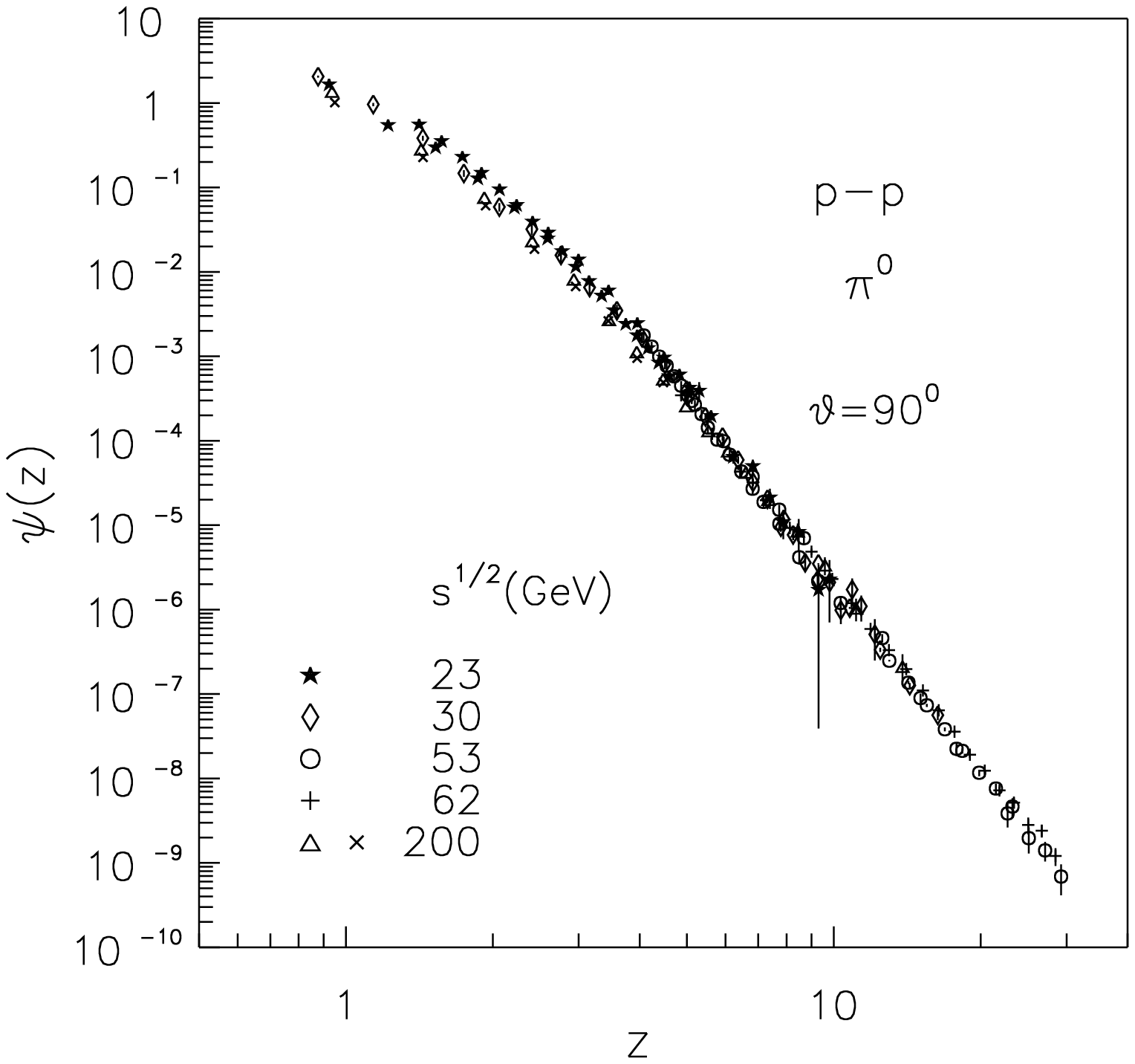}
\hspace*{5mm}
\includegraphics[width=10pc,height=10pc]{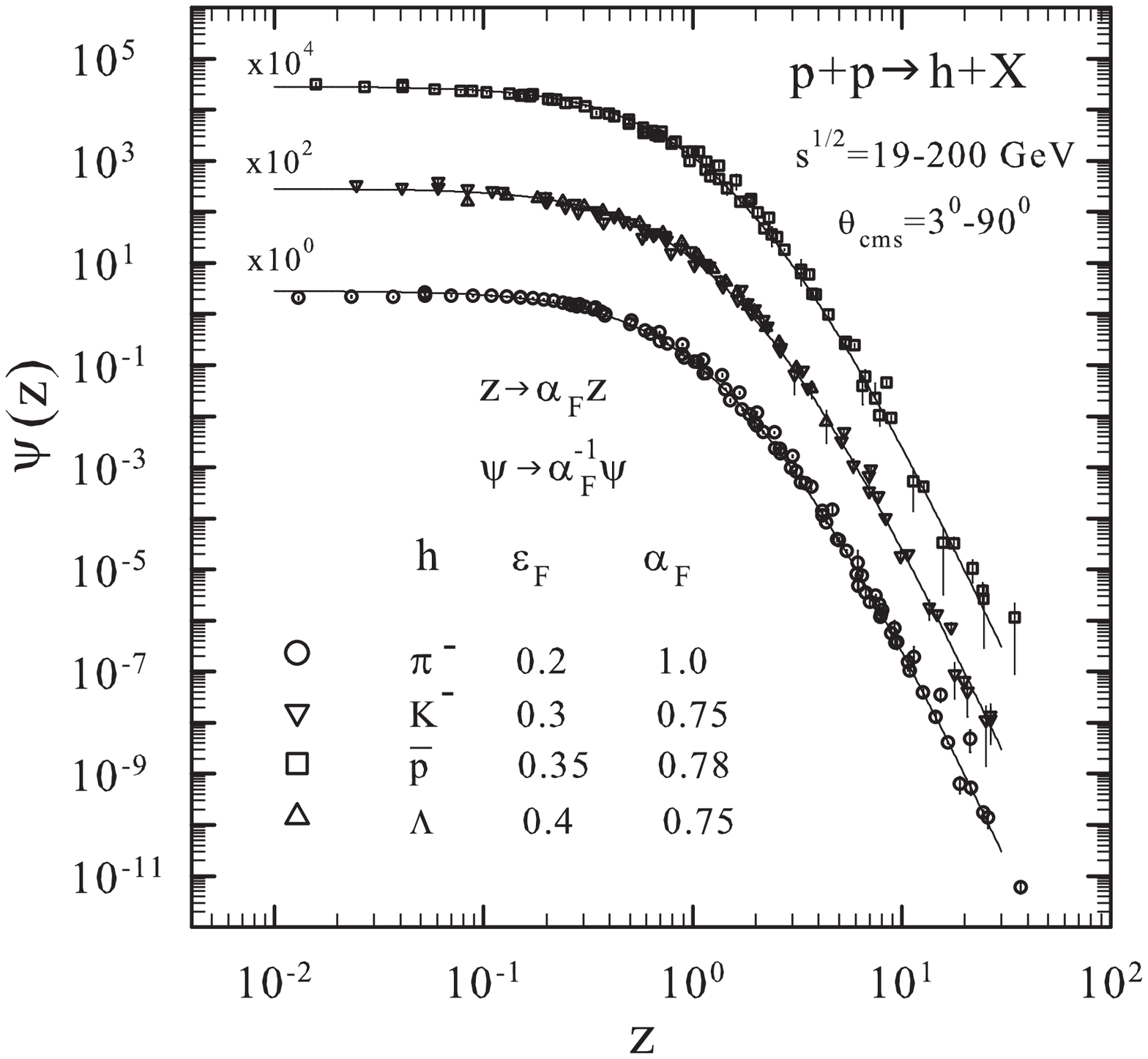}
\hspace*{23mm} a) \hspace{52mm}  b)  \hspace{52mm}  c)
\caption{\label{label} The data $z$-presentation of inclusive
 spectra of jet (a) $\pi^0$ meson (b) and hadron (c)
 production in  $p+p$ and $\bar p+p$ collisions 
 at ISR, $\rm Sp\bar p S$ and RHIC energies. 
}
\end{figure}
As seen from the figure, similarity 
of the inclusive cross sections 
is valid in a wide range of kinematic variables.
The function $\Psi(z)$ exhibits a power behavior at high $z$.

Extension of the method for reactions with polarized
particles was tested with data obtained at RHIC. 
We have constructed the functions $\Psi_{++}$,  $\Psi_{+-}$ and $\Psi_{+0}$ in dependence 
on corresponding scaling variables to verify the hypothesis (\ref{eq:19}).
Figure 2 demonstrates coincidence of the ratios of spin-dependent 
and spin-independent functions in dependence on $z$. The symbols shown 
in Fig. 2(a) correspond to the inclusive cross section \cite{8} 
and the spin asymmetry $A_{LL}$ \cite{3}
of jet production in proton-proton collisions 
measured by the STAR collaboration at $\sqrt s = 200$~GeV.
The data allows us to estimate the spin-dependent 
fractal dimension of proton from the process.

\begin{figure}[h]
\includegraphics[width=10pc,height=10pc]{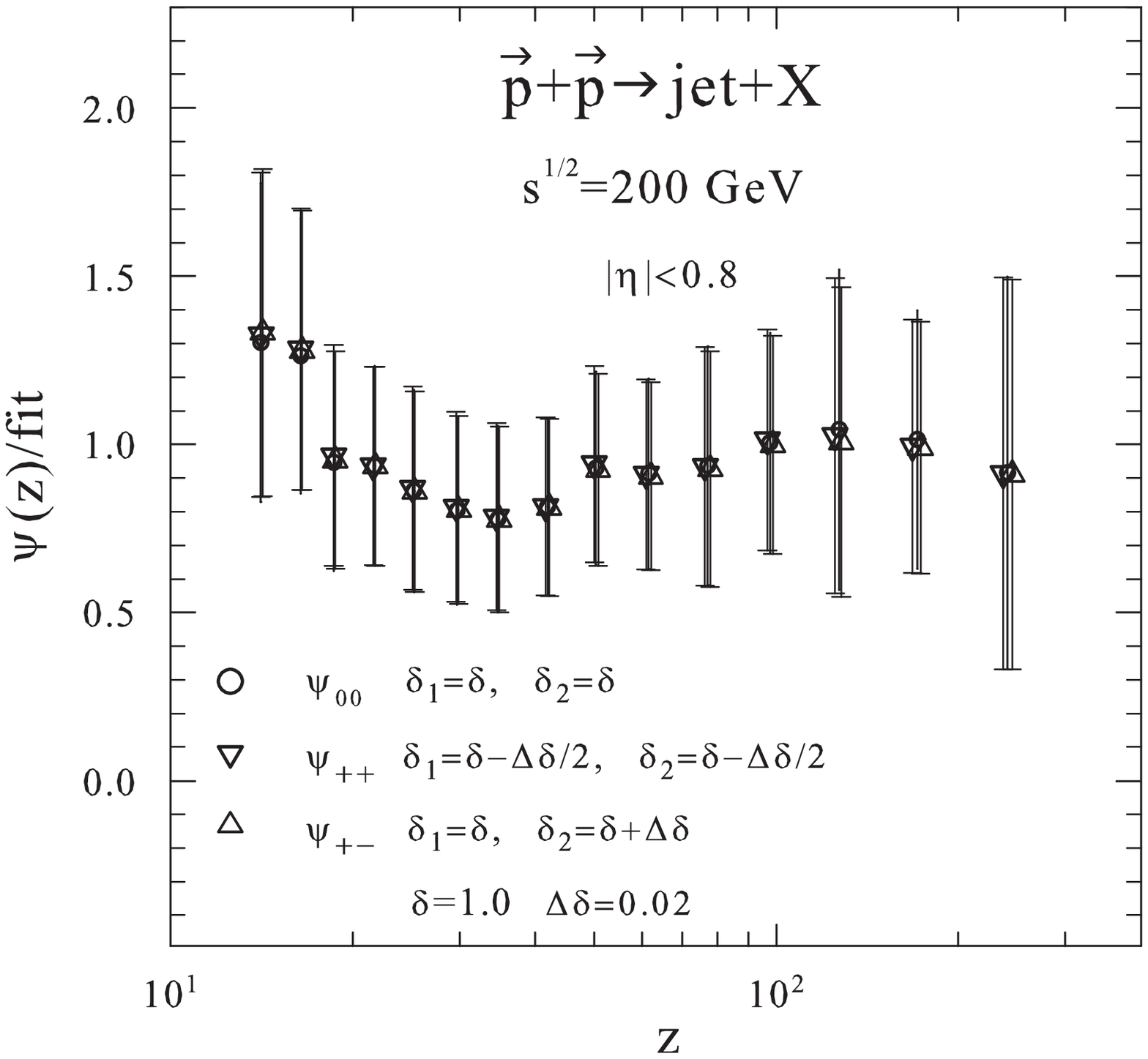}
\hspace*{5mm}
\includegraphics[width=10pc,height=10pc]{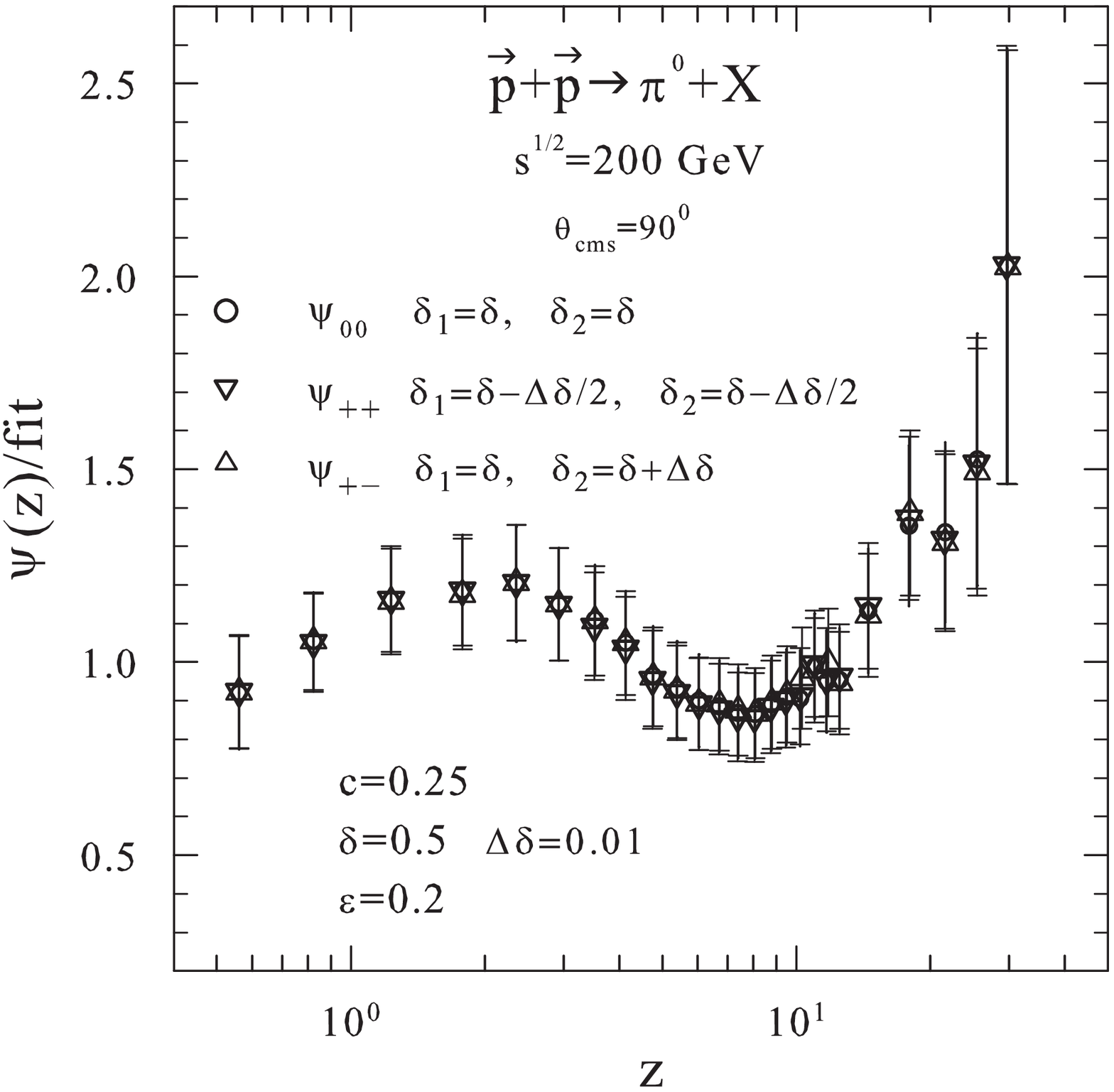}
\hspace*{5mm}
\includegraphics[width=10pc,height=10pc]{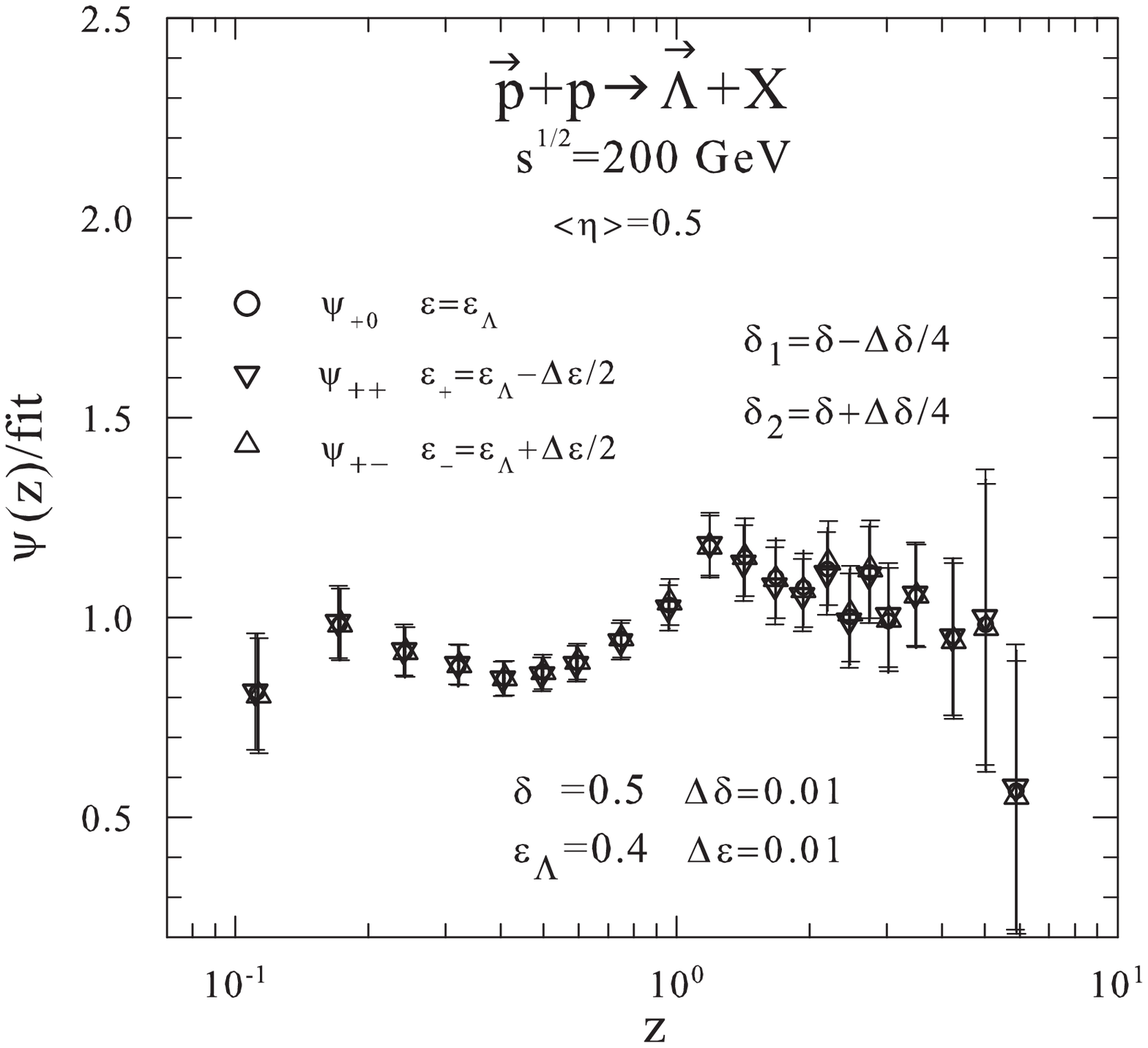}
\hspace*{23mm} a) \hspace{52mm}  b)  \hspace{52mm}  c)
\caption{\label{label} 
The ratios of scaled spin-dependent and spin-independent  functions 
for $ \overrightarrow{\!\!p}+ \overrightarrow{\!\!p} \rightarrow jet +X$ (a)
$ \overrightarrow{\!\!p}+ \overrightarrow{\!\!p} \rightarrow \pi^0 +X$ (b) and 
$\overrightarrow{\!\!p}+ p\rightarrow \overrightarrow{\!\!\Lambda} +X$ (c)
processes at $\sqrt s = 200$ GeV.}
\end{figure}
Figure 2(b) demonstrates 
the scaled spin-dependent functions $\Psi(z)/$fit  for the reaction   
$\overrightarrow{\!\!p}+\overrightarrow{\!\!p} \rightarrow \pi^{0}+X$. 
The double-longitudinal spin asymmetry $A_{LL}$
of $\pi^0$ mesons measured by the PHENIX collaboration 
at $\sqrt s  = 200$~GeV and  $\theta_{cms}=90^0$ \cite{7}
and corresponding unpolarized cross section were used  
in the analysis.
The correction to the fractal dimension $\delta$ 
of unpolarized proton is found to be  $\Delta \delta= 0.01$.
Figure 2(c) shows ratios  
of the spin-dependent and spin-independent functions 
for the process
$\overrightarrow{\!\!p}+p \rightarrow \overrightarrow{\!\!\Lambda}+X$.
The comparison exploits data on the 
longitudinal spin transfer coefficient $D_{LL}$ 
measured by the STAR collaboration at $\sqrt s  = 200$~GeV and $<\eta>= 0.5$ \cite{6}.
Using the value of $\Delta \delta =0.01$ obtained from the analysis 
of  $\pi^0$-meson production, the spin-dependent correction 
to $\epsilon_{\Lambda}$ is found to be $\Delta \epsilon_{\Lambda}=0.01$.

Performed analysis justifies application of the $z$-scaling approach to study 
the scale properties of spin structure in hadron interactions with polarized particles.
The coincidence of functional ratios shown in Fig. 2 
indicates self-similarity of spin-dependent processes expressed by (\ref{eq:19}). 
This 
condition 
was used to estimate corrections $\Delta{ \delta}$ and $\Delta {\epsilon_{\Lambda}}$
to the fractal dimensions $\delta$ and $\epsilon_{\Lambda}$. 
The suggested procedure of data analysis 
is applicable to a wide class of polarization reactions. 
A systematic phenomenological investigation of the processes with polarizations
based on the self-similarity principle
would contribute to further development of theory and understanding 
of spin as one of the most important and basic property of particles.

\smallskip


\medskip
\begin{thebibliography}{9}
\bibitem{1} 
Zborovsk\'{y} I and Tokarev M V 
2007 {\it Phys. Rev.} D {\bf 75} 094008

\bibitem{2} 
Zborovsk\'{y} I and Tokarev M V 
2009 {\it  Int. J. Mod. Phys.}  A {\bf 24} 1417

\bibitem{4}
Tokarev M V, Zborovsk\'{y} I and Aparin A A 2015 {\it Part. Nucl. Lett.} {\bf 12} 81  

\bibitem{5}
Tokarev M V and Zborovsk\'{y} I 2015 {\it Part. Nucl. Lett.} {\bf 12} 214



\bibitem{8}
 Li X {\it Jet measurements in polarized p+p collisions at STAR at RHIC}
 ({\it Preprint} hep-ex/1506.06314)
 
 
\bibitem{3} 
Adamczyk L {\it et al}
2015 {\it Phys. Rev. Lett.} {\bf 115} 092002

 

\bibitem{7}
Arschenauer E C {\it et al} The RHIC Spin Program: Achievements and Future Opportunities
({\it Preprint } nucl-ex/1304.0079







\bibitem{6}
 Xu Q 2013 {\it Proc. XV Advanced Research Workshop on High Energy Spin Physics (DSPIN-13)}
(Dubna: JINR) p 333
  









\end{thebibliography}
\end{document}